\documentstyle[prl,aps]{revtex}

\tighten

\begin{document}

\draft
\title{
Robustness of Wave Functions of Interacting Many Bosons
in a Leaky Box
}
\author{
Akira Shimizu\cite{email} and Takayuki Miyadera
}
\address{
Department of Basic Science, University of Tokyo, 
Komaba, Meguro-ku, Tokyo 153-8902, Japan
}

\date{Received 16 February 2000}
\maketitle
\begin{abstract}
We study the robustness,
against the leakage of bosons, 
of wave functions of 
interacting many bosons confined in a finite box,
by deriving and analyzing 
a general equation of motion for the reduced density 
operator. 
We identify a robust wave function
that remains a pure state, whereas
other wave functions, 
such as 
the Bogoliubov's ground state
and the ground state with a fixed number of bosons,
evolve into mixed states.
Although these states all have the off-diagonal long-range order, 
and almost the same energy densities,
we argue that only the robust state is realized as a macroscopic
quantum state.
\end{abstract}
\pacs{PACS numbers: 
3.75.Fi, 03.65.Bz, 
05.70.Ln, 64.60.-i
}




When a quantum system is subject to perturbations 
from its environment, 
most wave functions decohere, 
and only an exceptional wave function(s) remains pure.
For quantum systems with a single degree of freedom ($f=1$),
this robust wave function is a coherent state \cite{QO,ZHP}.
For example, when 
a coherent state $|\alpha \rangle$ of single-mode photons
passes through an absorptive medium, 
the final state is also 
a coherent state $|\alpha' \rangle$, which 
was attenuated ($|\alpha'| < |\alpha|$) by the absorption \cite{QO}.
It was also argued 
for a $f=1$ system
 that coherent states produce the least entropy in the environment, 
thus being stable \cite{ZHP}.
Since these conclusions are based on 
analyses on $f=1$ systems,
a natural question is:
Are they applicable to macroscopic systems, i.e., to 
$f \gg 1$ interacting systems?
Moreover, we must identify which coherent states are robust,
because there are many choices of the coordinate 
(among many degrees of freedom)
by which a coherent state is defined. 
Furthermore, 
for massive bosons the superselection rule (SSR) 
forbids superpositions of states with different numbers of bosons.
Hence, we must clarify how coherent states can be compatible with 
the SSR.
The purpose of this {\it Letter} is to 
answer these questions for 
condensates of interacting many bosons, 
which (or, equivalents of which) are observed 
in many physical systems such as liquid He \cite{bec93}, 
quantum Hall systems \cite{QH}, 
excitons \cite{exciton},
and trapped atoms \cite{atom}.
We also discuss the symmetry breaking in view 
of the robustness.


We consider many bosons which interact with each other repulsively.
We assume that the bosons are confined in a large, but finite 
box of volume $V$, 
which is placed in a huge room of volume ${\cal V} \gg V$, 
which we call the environment.
Suppose that the potential of the walls of the box is not 
high enough, 
so that the box and the environment exchange bosons via tunneling 
processes at a small rate (flux) $J$.
Let $t_{\rm eq}$ denote the time scale after which 
the total system, the box plus the environment, reaches 
the equilibrium state.
We are not interested in the time region $t \geq t_{\rm eq}$ 
because the equilibrium state
is just a uniform state that is determined solely
by the initial state of the environment
(because ${\cal V} \gg V$).
We therefore examine the {\em transient region} for which 
$ 
t \ll t_{eq},
$ 
in order to discuss the robustness of an initial state $| \phi(0) \rangle$, 
which is prepared at $t=0$, 
of the box.
Depending on the choice of
the initial state $| \phi_{\rm E}(0) \rangle_{\rm E}$ of the environment, 
the box state may be affected either drastically or moderately.
For example, a moderate situation is that 
$| \phi_{\rm E}(0) \rangle_{\rm E}$ 
has the same density $n$ of bosons as $| \phi(0) \rangle$.
In such a case, $n$ of the box will be kept constant for all $t$.
To discuss the robustness, however, 
we consider the severest  
situation where the environment is initially in the vacuum state 
$| 0 \rangle_{\rm E}$ of bosons, 
so that bosons escape from the box continuously.
If a box state is robust in 
this severest case, it would also be robust in other cases. 
Hence, the total wave function at $t=0$ is
$ 
| \Phi(0) \rangle_{\rm total}
=
| \phi(0) \rangle
\otimes
| 0 \rangle_{\rm E}.
$ 
%
%
We decompose (the ${\bf r}$ dependence of) the 
boson field that is defined on $V + {\cal V}$ as
$ 
\hat \psi_{\rm total} ({\bf r})
=
\hat \psi({\bf r})
+
\hat \psi_{\rm E} ({\bf r}).
$ 
Here, $\hat \psi({\bf r})$ localizes in the box, whereas
the low-energy component of $\hat \psi_{\rm E} ({\bf r})$ 
localizes in the environment \cite{psi_psiE}.
Accordingly, the Hamiltonian of the total system is 
decomposed as
$ 
\hat H_{\rm total}
=
\hat H + \hat H_{\rm E} + \hat H_{\rm SE}.
$ 
Here, $\hat H$ ($\hat H_{\rm E}$) is a function of 
$\hat \psi$ ($\hat \psi_{\rm E}$) only, describing 
interacting bosons in the box (environment).
On the other hand, 
$\hat H_{\rm SE}$ includes both $\hat \psi$ and $\hat \psi_{\rm E}$,
describing the $\hat \psi$-$\hat \psi_{\rm E}$ interaction.
If the leakage flux $J$ is small, 
the probability of finding two or more bosons simultaneously in 
a wall of the box is negligible, and thus 
the dominant term of $\hat H_{\rm SE}$ takes 
the following form:
\begin{equation}
\hat H_{\rm SE}
=
\lambda
\int d^3 {\bf r} \ 
\hat \psi_{\rm E}^\dagger({\bf r}) w({\bf r}) \hat \psi({\bf r})
+ {\rm H.c.}
\label{HSE}\end{equation}
Here, $w({\bf r})$ represents the shapes of the walls
($w \sim 1$ in the walls, $w=0$ in other regions),
whose potential height is characterized by 
a parameter $\lambda$.
Details of $w$
are irrelevant because 
they are all absorbed in the 
value of $j$, Eq.\ (\ref{j}).
In the time region of interest ($t \ll t_{eq}$), 
the $\hat \psi_{\rm E}$-$\hat \psi_{\rm E}$ interaction 
should be unimportant because $n$ of the environment 
remains zero.
On the other hand, 
we must treat the $\hat \psi$-$\hat \psi$ interaction
appropriately. 
For this purpose, 
we use the 
decomposition formula for $\hat \psi$ \cite{LP,SI,unpublished}:
\begin{equation}
\hat \psi
=
\hat \Xi + \hat \psi',
\label{decomposition}\end{equation}
where $\hat \Xi$ is an operator satisfying
\begin{equation}
\hat \Xi | N, {\rm G} \rangle
=
\sqrt{N} \xi | N-1, {\rm G} \rangle,
\label{hatXi_SI}
\end{equation}
where
$| N, {\rm G} \rangle$ denotes the ground state 
that has {\em exactly} $N$ bosons\cite{LP,SI,GA}, 
which we call the number state of interacting bosons (NSIB),
and 
\begin{equation}
\xi 
\equiv
\langle N-1, {\rm G}| \hat \psi | N, {\rm G} \rangle/\sqrt{N}
\label{xi}\end{equation}
is a hallmark of the condensation:
$\sqrt{N} \xi = {\cal O}(1)$ for condensed states, whereas
$\sqrt{N} \xi = {\cal O}(1/\sqrt{V})$ for normal states 
\cite{LP,SI,r_dependence}.
We here consider the condensed states.
Since $\hat \psi$ alters $N$ exactly by 1, 
Eq.\ (\ref{xi}) means that 
$ 
{\langle N - \Delta N, {\rm G} |}  \hat \psi {| N, {\rm G} \rangle}
=
\sqrt{N} \xi \ \delta_{\Delta N, 1}
$ 
for all $\Delta N$ such that $|\Delta N| \ll N$.
It then follows from Eqs.\ (\ref{decomposition}) and 
(\ref{hatXi_SI}) that 
\begin{equation}
{\langle N - \Delta N, {\rm G} |}  \hat \psi' {| N, {\rm G} \rangle}
=
0
\quad
\mbox{(for $|\Delta N| \ll N$)}.
\label{me_psi_prime}\end{equation}
Namely, $\hat \psi'$ transforms $| N, {\rm G} \rangle$ into 
excited states.

For weakly-interacting bosons,
the explicit forms of the NSIB were given in
Refs.\ \cite{SI,GA}, and that of 
$\hat \Xi$ was given in \cite{SI}.
Because of the boson-boson interaction,
they are rather complicated functions of 
bare operators $\hat a_{\bf k}$:
$
| N, {\rm G} \rangle
=
(1/\sqrt{N!}) e^{i \hat G} (\hat a_0^\dagger)^N
| 0 \rangle
$
and
$ 
\hat \Xi
=
e^{i \varphi}
\sqrt{n_0/n V}
e^{i \hat G} \hat a_0 e^{-i \hat G}
$. 
Here, 
$
n_0
=
\langle
\hat N - \int d^3 {\bf r} \hat \psi^{\prime \dagger} \hat \psi' 
\rangle/V
$,
$
\hat G 
\equiv
(-i/2 n V) \hat a_0^\dagger \hat a_0^\dagger 
\sum_{{\bf q} \neq {\bf 0}} y_q \hat a_{\bf q} \hat a_{-{\bf q}}
+ {\rm H.c.}
$,
$\varphi$ is an arbitrary phase, 
and $y_q$ is given in Ref.\ \cite{SI}.
Using these expressions, we can show that 
\cite{unpublished}
\begin{equation}
[\hat \Xi, \hat \Xi^\dagger]
, \
[\hat \Xi, \hat \psi']
, \
[\hat \Xi, \hat \psi^{\prime \dagger}]
\ = {\cal O}(1/V).
\label{com_rel}\end{equation}
Lifshitz and Pitaevskii (LP) \cite{LP}
claimed that Eq.\ (\ref{com_rel}) is applicable 
even when the interaction is stronger.
Their discussion is somewhat controversial because 
LP started from, instead of Eq.\ (\ref{hatXi_SI}), 
the assumption that 
$\hat \Xi$ could be defined by
$ 
\hat \Xi | N, \nu \rangle
=
\Xi | N-1, \nu \rangle,
$ 
where 
$| N, \nu \rangle$ denotes {\it any} 
eigenstate 
that has exactly $N$ bosons.
However, we note that 
for weakly-interacting bosons
we have not used this assumption
in the derivation of Eq.\ (\ref{com_rel}).
We thus expect that 
Eq.\ (\ref{com_rel}) also holds
for bosons with stronger interaction,
even if LP's assumption was too strong.
If this is the case, 
the following results are applicable 
not only to weakly interacting bosons
but also to bosons with stronger interaction,
because the results will be derived only from 
Eqs.\ (\ref{HSE})-(\ref{com_rel}).

Since we are studying the robustness against {\em weak}
perturbations, we assume that $\lambda$ is small,
so that $J$ is very small.
In this case, 
we have to consider transitions only among
$| N, {\rm G} \rangle$'s with different $N$'s
(i.e., we can neglect transitions to
excited states).
Hence, the reduced density operator $\hat \rho$ can be
generally written as
\begin{equation}
\hat \rho(t)
=
\sum_{N,M}
\rho_{NM}(t)
| N, {\rm G} \rangle \langle M, {\rm G}|.
\end{equation}
It seems almost obvious that quantum coherence 
between $| N, {\rm G} \rangle$ and $| M, {\rm G} \rangle$ with large 
$|N-M|$ would be destroyed by the interaction with the environment.
We therefore study the most interesting case where 
$\rho_{NM}$ is localized in the $N$-$M$ plane in such a way that 
$
\sqrt{\langle \delta N^2 \rangle}
\ll
\langle N \rangle
$.
If this relation is satisfied at $t=0$, 
it is also satisfied for all $t \ll t_{\rm eq}$.
We also assume that ${\cal V}$ is large enough, 
so that the boson density in the environment
($=[\langle N(0) \rangle - \langle N(t) \rangle]/{\cal V}$)
is negligibly small for all $t \ll t_{\rm eq}$. 
Under these conditions, 
we can calculate the time evolution of $\rho_{NM}(t)$,
using Eqs.\ (\ref{HSE})-(\ref{com_rel}),
as\cite{unpublished,early}
\begin{eqnarray}
&&
\rho_{NM}(t + \Delta t)
=
e^{-i (N-M) \mu(n(t)) \Delta t / \hbar}
\nonumber\\
&& \quad \times
\left[
\rho_{NM}(t) 
(1- (N+M) j(n(t)) \Delta t/2)
\right. 
\label{master}\\
&& \quad + \left. 
\rho_{N+1,M+1}(t) \sqrt{(N+1)(M+1)} j(n(t)) \Delta t
\right]
+
{\cal O}(\lambda^4),
\nonumber
\end{eqnarray}
for a finite time interval $\Delta t$ that satisfies 
$
\hbar / E_c \lesssim \Delta t < 1/\langle N \rangle j(n)
$, 
where $E_c$ is the energy scale over which 
the matrix elements of $\hat H_{\rm SE}$ 
are non-negligible.
Here, $n \equiv \langle N \rangle/V$, 
and $\mu(n)$  
($>0$ for a condensate of interacting bosons \cite{bec93,LP,SI})
denotes the chemical potential
of bosons in the box.
Furthermore, 
\begin{equation}
j(n)
=
K
\frac{2 \pi}{\hbar}
\frac{n_0}{n}
|\lambda|^2
\frac{v^2}{V}
D(\mu(n)),
\label{j}\end{equation}
where
$D(\mu)$ is the density of states per unit volume of the 
environment at energy $\mu$, 
$v$ is the total volume of the walls of the box,
and $K$ is a constant of order unity.
Both $\mu$ and $j$ depend on $\langle N \rangle$ 
through $n$, but this dependence is very weak because
a change of $\langle N \rangle$ by 1 only causes the 
change of $n$ by $1/V$.
Note that our basic equation
 (\ref{master}) has only two parameters, $\mu$ and $j$.
Namely, all model-dependent parameters
(details of $\hat H$, $\hat H_{\rm E}$ and $\hat H_{\rm SE}$) 
are absorbed in these two parameters.
Therefore, the following results are general and model independent.

Using Eq.\ (\ref{master}), we first calculate 
the time evolutions of the expectation value
 $\langle N \rangle$ 
and the fluctuation 
 $\langle \delta N^2 \rangle$ 
of the number $N$ of bosons in the box.
We find 
\begin{equation}
\frac{\rm d}{{\rm d} t}
\langle N \rangle
=
-  j(n) \langle N \rangle.
\label{dNdt}\end{equation}
Hence, $\langle N \rangle$ decreases gradually because of 
the leakage flux $J = j(n) \langle N \rangle$.
For $\langle \delta N^2 \rangle$, on the other hand, we find 
\begin{equation}
\frac{\rm d}{{\rm d} t} F
=
j(n) [ 1 - F],
\label{dFdt}\end{equation}
where 
$ 
F
\equiv
\langle \delta N^2 \rangle /\langle N \rangle
$ 
is the ``Fano factor'' \cite{QO}.
It is seen that
a robust state must have $F=1$, whereas
any states with $F \neq 1$ are fragile in the sense that 
their $F$ evolves with time, approaching unity.
For example, the ground-state wave function in the 
Bogoliubov approximation, 
$| {\rm Bog}, {\rm G} \rangle$, 
has $F>1$ \cite{SI}.
Hence, it is fragile.
The ground-state wave function with 
a fixed number of bosons, 
$| N, {\rm G} \rangle$, 
is also fragile because $F=0$.

Since the evaluation of $F$ is easy,
$F$ is a convenient tool for the investigation of the robustness.
However, 
since $F$ is only related to the diagonal elements of $\hat \rho$, 
it does not distinguish between pure and mixed states.
Therefore, 
we now solve the basic equation (\ref{master})
for various initial states
to investigate the robustness of the wave functions in more detail.
When the initial state is a pure state of the NSIB, i.e., 
$\hat \rho(0)=
| N, {\rm G} \rangle \langle N, {\rm G}|$,
then 
$\rho_{N M}$ after a short interval $\Delta t$ is evaluated as
$
\rho_{NN}(\Delta t)
=
(1- N j(n(t)) \Delta t)
$, 
$
\rho_{N-1,N-1}(\Delta t)
=
N j(n(t)) \Delta t
$,
and other elements are zero.
Therefore, $\hat \rho$ becomes a 
classical mixture of $| N, {\rm G} \rangle$ and $| N-1, {\rm G} \rangle$ at $t=\Delta t$,
in consistency with the above result that states with $F=0$ are fragile.
By evaluating the evolution at later times, we find that
$\hat \rho$ evolves toward a Poissonian mixture of 
$| N, {\rm G} \rangle$'s \cite{early}, consistent with 
$F \to 1$.
In a similar manner, 
we can show that 
the pure state of Bogoliubov's ground state
$\hat \rho(0)=| {\rm Bog}, {\rm G} \rangle \langle {\rm Bog}, {\rm G} |$,
which has $F>1$,
also evolves into a mixed state.
We can also show that 
the number-phase squeezed state of interacting bosons (NPIB), 
which was found in Ref.\ \cite{SI} as a 
number-phase minimum uncertainty state with
$0 < F < 1$, 
also evolves into a mixed state.
These examples show that 
$F$ is indeed a simple measure of the robustness:
A pure state with $F \neq 1$ is unlikely to remain pure.
Note, however, that 
a pure state with $F = 1$ is not necessarily robust.
For example, we can show that 
the coherent state of free bosons (CSFB) evolves into a mixed state, 
although it has $F=1$.
Hence, $F=1$ is only a {\it necessary condition} for the robustness.

Among many states with $F=1$, 
we have successfully found a very special state that is robust 
in the sense that it remains pure
when it is weakly perturbed by the environment.
The state is given by
\begin{equation}
\hat \rho(t)
=
| \alpha(t), {\rm G} \rangle
\langle \alpha(t), {\rm G} |.
\label{rho_CSIB}\end{equation}
Here, $\alpha(t)$ is a time-dependent complex number given by
\begin{equation}
\alpha(t)
=
e^{i \varphi(t)}
\sqrt{\langle N(t) \rangle},
\label{alpha_t}\end{equation}
where
$\langle N(t) \rangle$ is the solution of Eq.\ (\ref{dNdt}),
and
\begin{equation}
\varphi(t)
=
\varphi(0) - \frac{i}{\hbar} \int_0^t \mu(n(t)) {\rm d} t.
\end{equation}
Here, the initial phase $\varphi(0)$ is arbitrary, and 
$n(t) \equiv \langle N(t) \rangle / V$.
Furthermore, 
\begin{equation}
| \alpha, {\rm G} \rangle
\equiv
e^{- |\alpha|^2/2}
\sum_{M=0}^{\infty}
\frac{\alpha^M}{\sqrt{M!}}
|M, {\rm G} \rangle,
\label{CSIB}\end{equation}
which we call the coherent state of interacting bosons (CSIB).
It has the same form as the CSFB
except that $|M, {\rm G} \rangle$ is the NSIB.
Because of this difference, 
simple relations for the CSFB do not hold for the CSIB.
For example, 
$
\langle \alpha, {\rm G} | \hat \psi | \alpha, {\rm G} \rangle
\neq
\alpha/\sqrt{V}
$,
and, moreover, 
$
| \alpha, {\rm G} \rangle
$
is not an eigenstate of $\hat \psi$.
Nevertheless, 
$
\langle \alpha, {\rm G} | \hat N | \alpha, {\rm G} \rangle
=
\langle \alpha, {\rm G} | \delta \hat N^2 | \alpha, {\rm G} \rangle
=
|\alpha|^2
$,
hence $F=1$ exactly, 
as in the case of CSFB.
Since the NSIB has a complicated wave function, 
so does the CSIB.
[For weakly-interacting bosons, 
its explicit form was given in Ref.\ \cite{SI}.]
Although complicated, 
the wave function of the CSIB is 
robust against weak perturbations from the environment: 
It keeps the same form, 
whose parameter $\alpha(t)$ evolves slowly (except for the 
phase rotation), 
and remains a pure state, 
in contrast to other wave functions which 
soon evolve into mixed states.
In fact,
Eqs.\ (\ref{rho_CSIB})-(\ref{CSIB})
yield
\begin{eqnarray}
\rho_{NM}(t)
&=&
e^{- \langle N(t) \rangle}
\frac{e^{i (N-M) \varphi}}{\sqrt{N! M!}}
\langle N(t) \rangle^{\frac{N+M}{2}},
\\
\rho_{NM}(t + \Delta t)
&=&
e^{- \langle N(t) \rangle (1-j \Delta t)}
\frac{e^{i (N-M) (\varphi - \mu \Delta t / \hbar)}}{\sqrt{N! M!}}
\nonumber\\
& & \times
\langle N(t) \rangle^{\frac{N+M}{2}}
(1-j \Delta t)^{\frac{N+M}{2}},
\end{eqnarray}
which indeed satisfy Eq.\ (\ref{master}).

We now discuss the compatibility with the SSR, which 
might raise the objection that
the CSIB would not be realized because 
superpositions 
between states with different values of $N$ are forbidden
for massive bosons. 
To show that this intuitive objection is wrong, 
it is sufficient to give 
one counterexample. 
Suppose that there is another box, 
which also contains condensed bosons,
in the same room.
The total system consists of two boxes and 
the environment. 
According to the SSR, the wave function of the total system
$| \Phi \rangle_{\rm total}$ 
should be a superposition of states that have 
the same number of bosons,
$N_{\rm total} = N + N' +N_{\rm E} =$ fixed, 
where $N'$ denotes the number of bosons in the second box.
Consider the following state, which satisfies this constriction;
\begin{eqnarray}
| \Phi \rangle_{\rm total}
&=&
\sum_{N,N',\ell}
e^{- |\alpha|^2/2 - |\alpha'|^2/2}
\alpha^N {\alpha'}^{N'} C_{\ell} /\sqrt{N! N'!} 
\nonumber\\
&\times&
| N, {\rm G} \rangle
\otimes
| N', {\rm G} \rangle'
\otimes
| N_{\rm total} - N - N', \ell \rangle_{\rm E}.
\label{Psi_total}\end{eqnarray}
Here,
$\alpha = |\alpha| e^{i \varphi}$, 
$\alpha' = |\alpha'| e^{i \varphi'}$, 
and
$C_\ell$ is a complex number,
where
$\ell$ is a quantum number labeling states of the 
environment 
$| M, \ell \rangle_{\rm E}$ which has $M$ bosons.
Regarding the phases $\varphi$ and  $\varphi'$,
only the relative value  $\varphi -  \varphi' \equiv \theta$  
has a physical meaning. 
We thus take $\varphi'=0$ henceforth.
Equation (\ref{Psi_total}) yields
the reduced density operator of the first box as
$ 
\hat \rho
=
\sum_N
e^{- |\alpha|^2} (|\alpha|^{2 N} / N!)
| N, {\rm G} \rangle \langle N, {\rm G} |
$. 
It is easy to show that this is identical to
\begin{equation}
\hat \rho
=
\int_{-\pi}^{\pi}
\frac{d \theta}{2 \pi}
| |\alpha|e^{i \theta}, {\rm G} \rangle \langle |\alpha|e^{i \theta}, {\rm G} |.
\label{mixed_CSIB}\end{equation}
Although this $\hat \rho$ represents a mixed state of CSIB's,
we note that {\em it does not contain the maximum information}
on the state in the box, 
whereas the best density operator should have the maximum information.
The lacking information is 
that the phase relative to 
the condensate in the second box is $\varphi$.
Hence, the maximum information is 
Eq.\ (\ref{mixed_CSIB}) with the restriction $\theta=\varphi$.
This combined information is concisely expressed as
$ 
\hat \rho
=
| |\alpha|e^{i \varphi}, {\rm G} \rangle \langle |\alpha|e^{i \varphi}, {\rm G} |,
$ 
which agrees with Eq.\ (\ref{rho_CSIB}).
Namely, Eq.\ (\ref{rho_CSIB}) is better than Eq.\ (\ref{mixed_CSIB}) 
because it contains more information.
This example demonstrates that
Eq.\ (\ref{rho_CSIB}) can be compatible with the SSR
in realistic cases where the box exchanges bosons with the environment.
Only in the limiting case where
the box is completely closed,
should the SSR be crucial, and 
the NSIB would be realized
if the temperature $T \to 0$ \cite{finiteT}.

We have established that 
the CSIB is a robust pure state of interacting many bosons.
We finally discuss its implications.
The robustness of the present work
should not be confused with 
the ``stiffness of macroscopic wave functions'' \cite{bec93,QH},
which only referes to the stability of an order parameter
in a mean field approximation.
For example, $| {\rm Bog}, {\rm G} \rangle$ 
has the stiffness \cite{bec93,QH}, 
whereas it is fragile as we have shown.
The robustness is generalization of 
the robustness of coherent states 
of $f=1$ systems \cite{QO,ZHP}.
It is thus natural to expect for $f \gg 1$ systems that 
{\em some} coherent state would be robust.
However, it was not known 
{\em which} coherent state is robust:
there are many choices 
of the coordinate by which a coherent state is defined. 
Since Eq.\ (\ref{hatXi_SI}) yields
\begin{equation}
(\hat \Xi/\xi) | \alpha, {\rm G} \rangle
=
\alpha | \alpha, {\rm G} \rangle, 
\end{equation}
the present work has revealed that 
the robust coherent state is the one defined
by $\hat \Xi/\xi$.
In this sense, $\hat \Xi + \hat \Xi^\dagger$ is 
the ``natural coordinate'' of interacting many bosons.
The condensation of bosons are often characterized by 
the off-diagonal long-range order (ODLRO) that is defined by
$ 
\langle
\hat \psi^\dagger({\bf r}) \hat \psi({\bf r'})
\rangle
=
$
finite for
$
|{\bf r} - {\bf r'}| \sim V^{1/3}
$ \cite{bec93}. 
Using Eqs.\ (\ref{decomposition})-(\ref{me_psi_prime}) 
we can show that  
the CSIB, NSIB, NPIB, and the Bogoliubov's ground state
all have the ODLRO.
Hence, the present work has revealed that 
the ODLRO does not necessarily imply the robustness.
Furthermore, 
all of these states 
have almost the same energy densities,
i.e., the differences of $\langle \hat H \rangle/V$ are only
${\cal O}(1/V)$ for the same value of $\langle N \rangle$ \cite{energy}.
For example, if we let $E_{N,G}$ be the eigenenergy of the NSIB,
$\hat H |N,G \rangle = E_{N,G} |N,G \rangle$, 
we can then easily show from Eq.\ (\ref{CSIB}),
neglecting terms of ${\cal O}(1/V)$, that
$
\langle \alpha, {\rm G} | \hat H | \alpha, {\rm G} \rangle/V
=
E_{|\alpha|^2,G}/V
=
E_{\langle N \rangle,G}/V
$.
Therefore, 
the robustness of the CSIB is {\em not} due to 
an energy difference, but to natures of wave functions.
Since interactions with the environment are finite
in most physical systems,
we argue that only the robust state, CSIB, 
should be realized as a macroscopic pure state. 
Since the (relative) phase of the CSIB is almost definite \cite{SI},
the global gauge symmetry is then broken.
Although $V$ is finite, 
we are thus led to the symmetry breaking
by considering the robustness.
This suggests that quantum phase transitions may have 
more profound origins than singularities that are developed 
as $V \to \infty$.
A conventional trick to get symmetry breaking states 
for boson condensates 
is to introduce a symmetry breaking field $\eta$, 
which couples to $\hat \psi$ as
$
\hat H_\eta
=
\int d^3{\bf r} ( \eta^* \hat \psi + \eta \hat \psi^\dagger )
$.
However, 
$\eta$ is usually considered as an unphysical field \cite{bec93,gf,KT}, 
and
it was sometimes argued that symmetry breaking states 
were meaningless because they look against the SSR \cite{KT}.
In contrast, the present work gives a physical reasoning
for the symmetry breaking, assuming only physical interactions, and 
shows the compatibility with the SSR.

This work has been supported by the CREST program 
of Science and Technology Corporation of Japan.


\begin{references}

\bibitem[*]{email} Electronic address: shmz@ASone.c.u-tokyo.ac.jp

\bibitem{QO}
H.\ P.\ Yuen, Phys.\ Rev.\ A {\bf 13}, 2226 (1976);
M.\ Ueda, Quantum Optics {\bf 1}, 131 (1989).

\bibitem{ZHP}
W.\ H.\ Zurek, S.\ Habib and J.\ P.\ Paz, 
Phys.\ Rev.\ Lett.\ {\bf 70}, 1187 (1993).

\bibitem{bec93}
See, e.g., papers in 
A.\ Griffin {\it et al.} (eds.), 
Bose-Einstein Condensation (Cambridge, New York, 1995).

\bibitem{QH}
S.\ M. Girvin and A.\ H.\ MacDonald, 
Phys.\ Rev.\ Lett.\ {\bf 58}, 1252 (1987);
S.\ C.\ Zhang, H.\ Hanson and S. Kivelson, 
Phys.\ Rev.\ Lett.\ {\bf 62}, 82 (1989);
Z.\ F.\ Ezawa, M.\ Hotta and A.\ Iwazaki,
Phys.\ Rev.\ B {\bf 46}, 7765 (1992).

\bibitem{exciton}
E. Fortin, S. Fafard, and A. Mysyrowicz,
Phys. Rev. Lett. {\bf 70}, 3861 (1993);
%
J. P. Wolfe, J. L. Lin and D. W. Snoke, 
p.\ 281 of Ref. \cite{bec93};
%
L.V. Butov {\it et al.}, 
{\it ibid}, {\bf 73}, 304 (1994).


\bibitem{atom}
M.\ H.\ Anderson {\it et al.},
Science {\bf 269} (1995) 198;
C.\ C.\ Bradley {\it et al.},
Phys. Rev. Lett. {\bf 75} (1995) 1687;
K.\ B.\ Davis {\it et al.},
Phys.\ Rev.\ Lett.\ {\bf 75} (1995) 3969.

\bibitem{psi_psiE}
Let $\varphi_\nu({\bf r})$ and
 $\varphi_\eta^{\rm E}({\bf r})$ be the solutions of 
the single-body Schr\"odinger equation on $V+ {\cal V}$ 
when the potential of the walls of the box is high.
Here, $\varphi_\nu$'s localize in the box, whereas
 $\varphi_\eta^{\rm E}$'s denote other solutions.
At low-energy, $\varphi_\eta^{\rm E}$'s localize in the environment.
Since $\{ \varphi_\nu, \varphi_\eta^{\rm E} \}$ form a complete set
of functions of ${\bf r}$, we can expand 
$
\hat \psi_{\rm total} 
$
as 
$
\hat \psi_{\rm total} ({\bf r})
=
\sum_\nu \hat a_\nu \varphi_\nu({\bf r})
+
\sum_\eta \hat a_\eta^{\rm E} \varphi_\eta^{\rm E}({\bf r})
\equiv
\hat \psi({\bf r})
+
\hat \psi_{\rm E} ({\bf r})
$.

\bibitem{LP} 
E.\ M.\ Lifshitz and L.\ P.\ Pitaevskii,
{\it Statistical Physics Part II} 
(Pergamon, New York, 1980), sec.\ 26.

\bibitem{SI}
A.\ Shimizu and J.\ Inoue, 
Phys.\ Rev.\ A {\bf 60}, 3204 (1999).

\bibitem{unpublished}
A.\ Shimizu and T.\ Miyadera, unpublished.

\bibitem{GA}
M.\ Girardeau and R.\ Arnowitt,
Phys.\ Rev.\ {\bf 113}, 755 (1959).

\bibitem{r_dependence}
For bosons confined in a box, 
$\xi$ generally depends on the position ${\bf r}$.
We here neglect this ${\bf r}$ dependence, because
it does not alter the conclusions of the present paper.

\bibitem{early}
An early time stage $t \ll 1/j(n)$
of this evolution 
was evaluated in Ref.\ \cite{SI}
for the special case where
$\hat \rho(0) = |N, {\rm G} \rangle \langle N, {\rm G}|$
and the boson-boson interaction is weak.

\bibitem{finiteT}
As $T$ is increased, 
an inhomogeneous state seems favorable 
which behaves locally as a CSIB, 
because an energy increase by the inhomogeneity
can be canceled by an increase of the entropy, 
leading to a minimum free energy. 

\bibitem{energy}
This agrees with the general theorems of Refs.\ 
\cite{gf,KT,rivers}.
This should not be confused with the result of 
the mean field approximation, according to which symmetry breaking ground states 
have the lowest energy.

\bibitem{gf} N.\ Goldenfeld, 
{\it Lectures on Phase Transitions and the Renormalization
Group}
(Addison-Wesley, New York, 1992).

\bibitem{KT}
T.\ Koma and H.\ Tasaki,
cond-mat/9708132, which is a revised version of
J.\ Stat.\ Phys.\ {\bf 76}, 745 (1994).

\bibitem{rivers}
R.\ J.\ Rivers, 
{\it Path integral methods in quantum field theory}
(Cambridge, 1987)
Chapter 13.




\end{references}
\end{document}